\documentstyle[12pt,epsf]{article}

\begin{document}

\hfill \vbox{\hbox{FTUAM-99-39} 
             \hbox{INLO-PUB-20/99}}
\begin{center}{\Large\bf Comments on the Instanton Size 
Distribution}\\[2cm] 

{\bf Margarita Garc\'{\i}a P\'erez} \\
{\em Departamento de F\'{\i}sica Te\'orica, C-XI, 
Universidad Aut\'onoma de Madrid, Madrid 28049, Spain}\\[6mm]

{\bf Tam\'as G. Kov\'acs}
and {\bf Pierre van Baal}\\
{\em Instituut-Lorentz for Theoretical Physics, 
P.O.Box 9506, NL-2300 RA Leiden, The Netherlands}\\[10mm]

\end{center}

\section*{Abstract}

By studying the non-linear effects of overlapping instanton pairs we 
address difficulties in the identification of instanton distributions
when the average instanton size is comparable to the average distance. 
For the exact charge two solution, we study how its parametrisation relates 
to a description in terms of individual instantons. There exist two dual 
sets of parameters describing the same charge two instanton solution. This 
duality implies the existence of a minimal separation between two instantons. 
Conventionally used lattice instanton finder algorithms based on the 
assumption of diluteness tend to underestimate instanton sizes. Finally 
we numerically confirm this for realistic parameters of the instanton 
liquid. The effect is enhanced by parallel orientation in group space.

\vfill
\pagebreak

\section{Introduction}

In recent years the instanton liquid model has been very successful in 
describing the low energy properties of light hadrons \cite{Shuryak}. 
On the other hand, it seems quite unlikely that instantons can account 
for confinement \cite{Noconfinement}. This is certainly the case if 
the instanton liquid is sufficiently dilute and the instanton size 
distribution falls off rapidly enough for large instantons. 
If the fall-off is slow, e.g.\ as $\rho^{-5}$ or $\rho^{-3}$, a linear 
term in the heavy quark potential due to instantons has been claimed 
\cite{Yesconfinement}, although these results are controversial. Still, 
a proper understanding of the tail of the instanton size distribution
may have physical significance.

Unfortunately very little is known about it. Lattice simulations 
are typically done on too small volumes and have too poor statistics 
to contain any precise information on the tail of the distribution. 
Moreover, even the question of what non-perturbative mechanism suppresses 
large instantons is still unanswered, although some recent attempts have 
been made \cite{Dual}. In this paper we would like to make some remarks 
on these issues by studying the most general charge two instanton solution. 
Our observations will be about the (possible mis)interpretations of the 
lattice data.

In lattice determinations \cite{Lat,I1,I2} of the instanton liquid parameters 
one usually starts with the assumption that the liquid is dilute enough, i.e. 
the individual pseudoparticles are far enough apart that they do not distort 
one another considerably. Only then does the ``instanton size'' have an 
unambiguous meaning. Treating the $Q=2$ case exactly can be thought of 
as the next order approximation when one takes into account the distorting 
effect of like charge nearest neighbour pairs.

To fix the notation, we start with briefly summarising the ADHM construction 
\cite{ADHM} of the general $Q=2$ solution~\cite{Christ}. Then we discuss the 
ambiguities arising in the identification between the ``physical'' parameters 
and the ones appearing in the ADHM construction, giving rise to two dual 
descriptions of the same physical configuration. This duality maps large 
separation between the constituents to small separation \cite{Dorey} and 
depends non-trivially on the relative gauge orientation of the two instantons. 
It has two important consequences for the identification of instantons from 
the charge density profile, most clearly seen in the two extreme cases. 
Namely parallel and perpendicular relative gauge orientation (the SU(2) 
invariant angle being 0 and $\pi/2$ respectively). If the orientation is 
perpendicular, the two instantons cannot be closer to each other than a 
minimal distance $\sqrt{2\rho_1\rho_2}$ set by their sizes $\rho_i$. If 
the relative orientation is parallel and the two constituents have the 
same size $\rho$, when they get close to each other, the charge density 
looks like the super-imposition of a small instanton of size proportional 
to the separation, plus another instanton of size $\sqrt{2}\rho$ right 
under the small one. If the relative orientation and sizes are not fine 
tuned, some combination of the above effects will take place.

The first effect controls how close two instantons can get to each other. 
The second one can potentially hide large instantons from instanton 
finding algorithms based on the diluteness assumption and thereby it 
can significantly distort the instanton size distribution. In the last 
part of the paper we present quantitative data on how these effects are 
manifested in realistic lattice situations. This is done by generating 
the charge density of instanton pairs, as given by the exact $Q=2$ 
solutions, and comparing the size distribution thus prescribed by the 
original data with that found by lattice instanton finder algorithms. 
We observe that the results depend very strongly on the relative 
orientation in group space, yielding a strong suppression of large 
instantons for parallel orientation. 

\section{Physical Parameters for $Q=2$ Instantons}

The fundamental objects describing the most general SU(2) charge
$Q$ instanton solution in the ADHM construction are the
$Q$-dimensional row vector $\lambda$ and the $Q\times Q$
symmetric matrix $B$, both having quaternionic elements \cite{ADHM}.
(A quaternion $x$ can be parametrised with 4 real numbers
as $x=x_\mu \sigma_\mu = x_0 + i x_k \tau_k$, where $\tau_k$
are the Pauli matrices.) The ADHM data can be conveniently 
summarised in a single $Q\times(Q+1)$ quaternionic matrix
$\Delta(x)$ which in the $Q=2$ case can be parametrised as
\begin{equation}
\Delta(x) = \left( \begin{array}{cc}
                \lambda_1   &    \lambda_2 \\
                y+z-x       &       u      \\
                u           &     y-z-x    
                   \end{array}
            \right).
\end{equation}
                    
$\Delta(x)$ describes a charge two self dual solution of the Yang-Mills
equations if and only if the matrix $\Delta(x)$ satisfies the ADHM
constraint that $\Delta^\dagger(x) \Delta(x)$ be real quaternionic (i.e. 
proportional to $\sigma_0$) and invertible.
Here $\lambda_{1,2}$, $y$, $z$, and $u$ are quaternionic
parameters, and $x=x_\mu \sigma_\mu$ denotes a space-time position. 
In terms of our parametrisation of the charge two case, the ADHM
constraint reads as
\begin{equation}
 \Lambda \equiv \frac{1}{2} \left( \lambda_2^\dagger \lambda_1 \,
              - \, \lambda_1^\dagger \lambda_2 \right) \; =
          \; z^\dagger u - u^\dagger z,
     \label{eq:Lambda}
\end{equation}
where the symbol $\Lambda$ is introduced to simplify the notation. 
This equation does not specify $u$ unambiguously; the most general 
solution can be written as \cite{Christ}
\begin{equation}
 u \; = \; \frac{z \Lambda}{2 |z|^2} + \alpha z,
     \label{eq:u}
\end{equation}
where $\alpha$ is an arbitrary real constant.

At this point it is instructive to compare this most general ADHM ansatz 
with the special case of the 't Hooft ansatz \cite{Hooft} in order to 
identify the physical parameters. This special case is obtained 
for parallel gauge orientations, $\lambda_1=\rho_1 \sigma_0$ and 
$\lambda_2=\rho_2 \sigma_0$ real and $u=0$, which solves the ADHM 
constraint, eq.\ (\ref{eq:Lambda}). We can identify $\rho_1$ and $\rho_2$ 
as the scale parameters (sizes), and $y \pm z$ as the locations of 
the two instantons. The action density of the solution, valid
for all choices of parameters \cite{Osborn}, 
\begin{equation}
 q(x) = -\partial_\mu^2\partial_\nu^2\log 
                        \det(\Delta^\dagger(x) \Delta(x))
     \label{eq:charge}
\end{equation}
indeed agrees in this case with the action density of the 't Hooft solution.
 
The most general charge two ADHM solution can be described by
the following set of free parameters: $\rho_{1,2}=|\lambda_{1,2}|$,
the scale parameters; $\lambda_1^\dagger \lambda_2/(|\lambda_1| 
|\lambda_2|) \in SU(2)$, the relative gauge orientation; and 
$y \pm z$ the location of the constituents. This gives a total 
number of 13 real parameters, in agreement with the general result
that the charge $Q$ solution has $8Q-3$ parameters. For $Q=2$
the conformal generalisation of the 't Hooft ansatz \cite{Hooft} 
also has 13 parameters, which can be related to the ADHM 
parametrisation~\cite{CFTG}.

The charge $Q$ ADHM ansatz is well known to have an $O(Q)$ 
symmetry acting on the parameters as
\begin{equation}
\lambda \rightarrow \lambda T^{-1}, \;\; 
B \rightarrow  T B T^{-1}, \;\; T \in O(Q).
\end{equation}
In the $Q=2$ case, using our parametrisation, the $SO(2)$ symmetry amounts to
\begin{equation}
\pmatrix{z\cr u}\rightarrow\pmatrix{\hphantom{-}\cos 2\psi&\sin 2\psi\cr
-\sin 2\psi&\cos 2\psi}\pmatrix{z\cr u},\quad
\pmatrix{\lambda_1\cr \lambda_2}\rightarrow\pmatrix{\hphantom{-}\cos \psi
&\sin \psi\cr -\sin \psi&\cos \psi}\pmatrix{\lambda_1\cr\lambda_2},
     \label{eq:O(2)}
\end{equation}
while $y$ is left unchanged. The $Z_2$ transformation that extends this
to the full $O(2)$ symmetry is generated by $T=\tau_1$, which interchanges
$\lambda_1$ with $\lambda_2$ and changes the sign of $z$. Different sets 
of parameters related by this $O(2)$ symmetry describe the same physical 
solution \cite{CFTG}
\begin{equation}
A_\mu(x)=\frac{1}{2}(1-\lambda F(x)\lambda^\dagger)^{-1}\partial_\nu
\left(\lambda\sigma^\dagger_{[\mu}\sigma^{\vphantom{\dagger}}_{\nu]}F(x)
\lambda^\dagger\right),\quad F^{-1}(x)\equiv\Delta^\dagger(x)\Delta(x),
\end{equation}
where $\lambda=(\lambda_1,\lambda_2)$. At first sight this seems to imply that
we have one less free parameter. This is, however, compensated by the fact
that, in general, solutions to the ADHM constraint with different values of 
$\alpha$ (eq.\ (\ref{eq:u})) result in physically different gauge field 
configurations, as one can easily be convinced of by computing 
$\det(\Delta^\dagger(x)\Delta(x))$. 

Ideally one would like to fix this ambiguity together with the $O(2)$ 
symmetry to have a one-to-one correspondence between the gauge inequivalent 
solutions and the 13 parameters describing them. Our choice of these 
parameters, which we call ``physical'', should be as close as possible to 
a superposition of two instantons. Looking at the situation of large 
separations ($|z|$ large), where the relative gauge orientation does not 
play a role, the action density should be the sum of two instantons of 
sizes $\rho_{1,2}$ located at $y \pm z$. This imposes $\alpha=0$, or 
equivalently $u_\mu z_\mu=0$. (The equivalence of the two conditions 
can be easily proved using eq.\ (\ref{eq:Lambda}) and 
$a^\dagger b + b^\dagger a = 2 a_\mu b_\mu$ which holds for any two 
quaternions $a$ and $b$.) Choosing this particular solution to the 
constraint ensures that when the orientation of the constituents is 
parallel, the ADHM solution coincides with the 't Hooft ansatz.

With this convention the identification of the ``physical'' parameters becomes 
almost unique. The prescription is that from each $O(2)$ orbit we have to 
choose the point that satisfies $u_\mu z_\mu = 0$. Generally\footnote{There 
are some degenerate cases, when $\Lambda=0$ or $|u|=|z|$.} there are 16 
such points on an $O(2)$ orbit describing the same gauge field configuration, 
as has been noted before \cite{Dorey}. They are generated by rotations 
(see eq.\ (\ref{eq:O(2)})) over multiples of $\pi/4$ and the $Z_2$ 
transformation, $\lambda_1\leftrightarrow\lambda_2$, $z\rightarrow -z$. 
Most of these do not affect the ``physical'' interpretation. But one
ambiguity remains for which we choose the representative described
by the $Z_2$ symmetry ($T=\tau_1$) combined with the rotation over
$\psi=\pi/4$, providing the relation
\begin{equation}
 \left( y, \, z, \, \lambda_1, \, \lambda_2, \,
                      u=\frac{z\Lambda}{2|z|^2} \right) \;\;
 \longrightarrow \;\;
 \left(y, \, u, \,
         \frac{\lambda_1+\lambda_2 }{\sqrt{2}}, \,
         \frac{\lambda_1-\lambda_2 }{\sqrt{2}}, \,
         z \right).
     \label{eq:duality}
\end{equation}
We note that if the distance of the two instantons is $2|z|$ in one 
description, it is proportional to $1/|z|$ in the ``dual'' description, 
as long as the relative gauge orientation is not parallel 
($\Lambda \neq 0$). Going beyond the issue of finding a unique 
parametrisation, the question now arises which of these two descriptions 
is the ``physical'' one. To answer this it is instructive to look at the 
charge density profile of a set of solutions with varying separations, 
keeping the other parameters fixed. 

\begin{figure}[!htb]
\begin{center}
\vskip 10mm
\leavevmode
\epsfxsize=140mm
\epsfbox{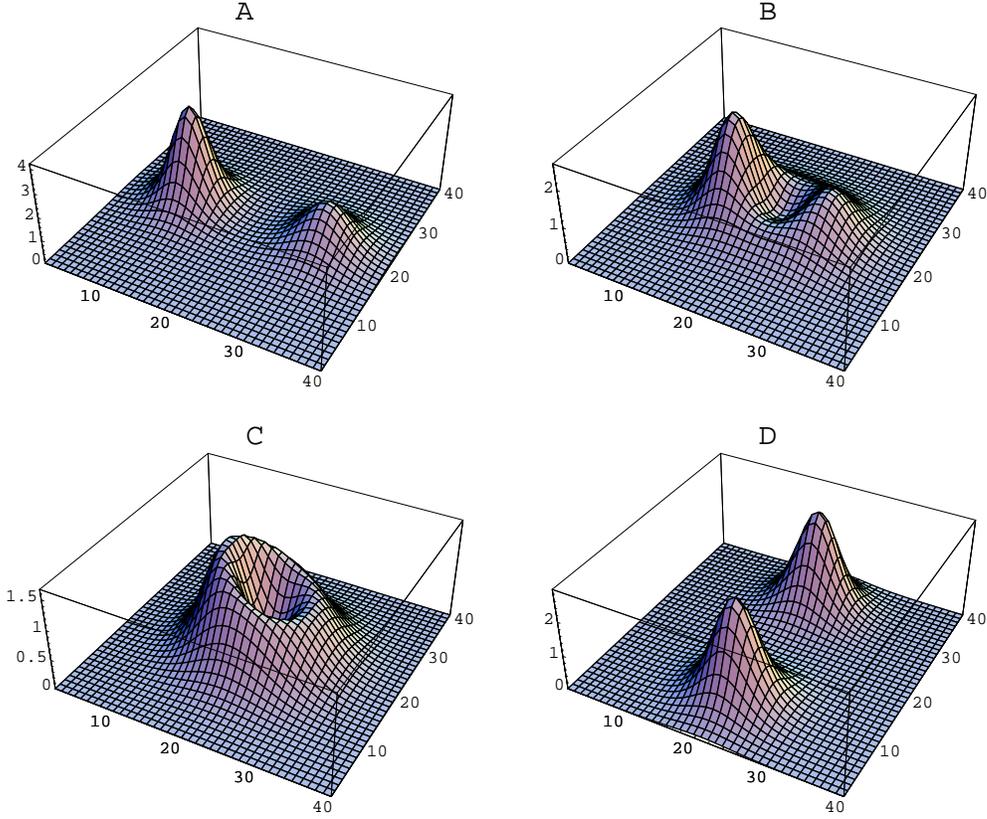}
\vskip 10mm
\end{center}
\caption{A pair of size $\rho_1=6.6$ and $\rho_2=8.3$ instantons
with perpendicular relative gauge orientations. The centres are
separated along the $x_0$ axis, the separation, $2|z|$ is 20.0 (A), 
13.3 (B), 10.5 (C), and 5.0 (D). The action density is shown 
in the (01) plane.}
\label{fig:scatter}
\end{figure}

In Fig.\ \ref{fig:scatter} we show such a sequence.
The parametrisation listed is in terms of the l.h.s.\ of 
eq.\ (\ref{eq:duality}). The scale parameters and relative orientation are 
described by $\lambda_1 = 6.6\sigma_0$, $\lambda_2=8.3\sigma_1$, and the 
separation between the constituents is {\em always} along the 0-axis. When 
they are far apart (A), corresponding to $2|z|= 20.0$, the solution indeed 
looks like a pair of instantons along the 0-axis. As the separation decreases
the two lumps merge together into an asymmetric ring (B-C) ($2|z|=13.3$ and 
10.5 respectively). For even smaller separation: $2|z|=5.0$, (D) in the 
figure, the two lumps separate again but now displaced along the 1-axis. 
Clearly, our parametrisation is not ``physical'' any more at this stage;
instead of two very close lumps separated along the 0-axis, we see two lumps 
farther apart but along the perpendicular 1-axis. On the other hand, we also 
have the dual parametrisation (r.h.s. of eq.\ (\ref{eq:duality})) at our 
disposal. Indeed, a short computation shows that in the dual description 
of (D) we have two instantons of the same scale parameter $\rho=7.5$
separated along the 1-axis at a distance of 22, evidently the correct 
``physical'' description.

The general picture emerging from this exercise is quite clear.
When $|z|$ is large, more precisely $|z|^2 \gg |\Lambda|$,
the original description is ``physical'', i.e. describing two
superposed instantons separated by a distance $2|z|$. When $|z|$ 
is small, however, the dual description is the more ``physical'' 
one. There is an intermediate region where the solution cannot 
be approximated by a combination of two instantons, here the 
question which parameter set to use for the physical description 
is ill defined. 

In the rest of this Section we discuss two simple consequences 
of this dual choice of parametrisation. The first one is that two 
instantons, as identified from maxima in the action density, can 
never get closer to each other than a minimal distance
\begin{equation}
  2|z|_{min}= \sqrt{2\rho_1 \rho_2 \sin \phi},
     \label{eq:zmin} 
\end{equation}
where $\phi$ is the invariant angle of the relative group 
orientation.  $|z|_{min}$ is defined by the property that the 
two dual descriptions have the same separation $2|z|_{min}$
between the constituents. If $|z|$ is chosen to be smaller
than this, the two instantons ``scatter off'' in the 
$z\Lambda$ direction and one has to switch to the dual
description. It is also interesting to
observe the very special case when $\rho_1=\rho_2=\rho$
and $\phi=\pi/2$. In this case, $2|z|_{min}=\sqrt{2}\rho$
gives a self-dual point where the two parametrisations
completely coincide and the charge density has an axial
symmetry in the $(z\Lambda,z)$ plane. For these parameters
the density profile is ring-like, not unlike the case of
monopoles \cite{Mon}, but apparently not noted before for
instantons. In particular, the solution does not degenerate
to one with topological charge $Q=1$, as conjectured in ref.
\cite{Dorey}. This {\em only} happens in the case of parallel
gauge orientation, for $\rho_1=\rho_2$ and $z=0$
(see eq.\ (\ref{eq:duality1}) below).

The discussion so far applies only to the $\Lambda \neq 0$ case.
Moreover, if $\phi \approx 0$, then the switching between
the two parametrisations occurs at very small $|z|$ (see eq.\
(\ref{eq:zmin})) and other interesting effects might
come into play when the two constituents are very close.
Let us now look at the extreme situation when $\phi=0$,
the relative orientation is parallel and consequently
$\Lambda=0$. In this case the two parametrisations are
\begin{equation}
 \left( y, \, z, \, \rho_1 \sigma_0, \, \rho_2 \sigma_0, \, 
                      0 \right) \;\;
 \longrightarrow \;\; 
 \left(y, \, 0, \, 
         \frac{\rho_1+\rho_2 }{\sqrt{2}} \sigma_0, \,
         \frac{\rho_1-\rho_2 }{\sqrt{2}} \sigma_0, \,
         z \right).
     \label{eq:duality1}
\end{equation}
In the limit $z\rightarrow 0$ the two descriptions become
equivalent, there is no way to choose between them. One can see
that two instantons of scale parameters $\rho_{1,2}$ ($\rho_1\sim\rho_2$)
on top of each other is equivalent to a small instanton of size
$|\rho_2-\rho_1|/\sqrt{2}$ on top of a larger one of size
$(\rho_2+\rho_1)/\sqrt{2}$. In this extreme case any lattice
instanton finder based on the diluteness assumption would
find only the former smaller instanton and nothing else.
Moreover due to the strongly non-linear character of the Yang-Mills
equations, when two instantons come close with parallel
orientation, both peaks become narrower and sharper and are thus identified
by any lattice algorithm as ``small'' instantons.
At best one misses one large instanton in the background;
at worst one misinterprets two large instantons as
two small ones. The effect is 
similar at $\phi \approx 0$ and can provide a mechanism 
to hide large instantons as some smooth background that 
remains unnoticed by lattice instanton finder algorithms.

\section{The Instanton Size Distribution}

We have seen how the identification of single instanton parameters becomes 
ambiguous when instantons overlap. We discussed what happens in the two 
extreme cases of parallel and perpendicular orientation in group space. 
In between, some combination of the two above described effects takes place. 
We expect that the exact way this affects the instanton size distributions
measured on the lattice will depend on the relative orientation of nearest 
neighbour pairs. 

\begin{figure}[!htb]
\begin{center}
\vskip 10mm
\leavevmode
\epsfxsize=120mm
\epsfbox{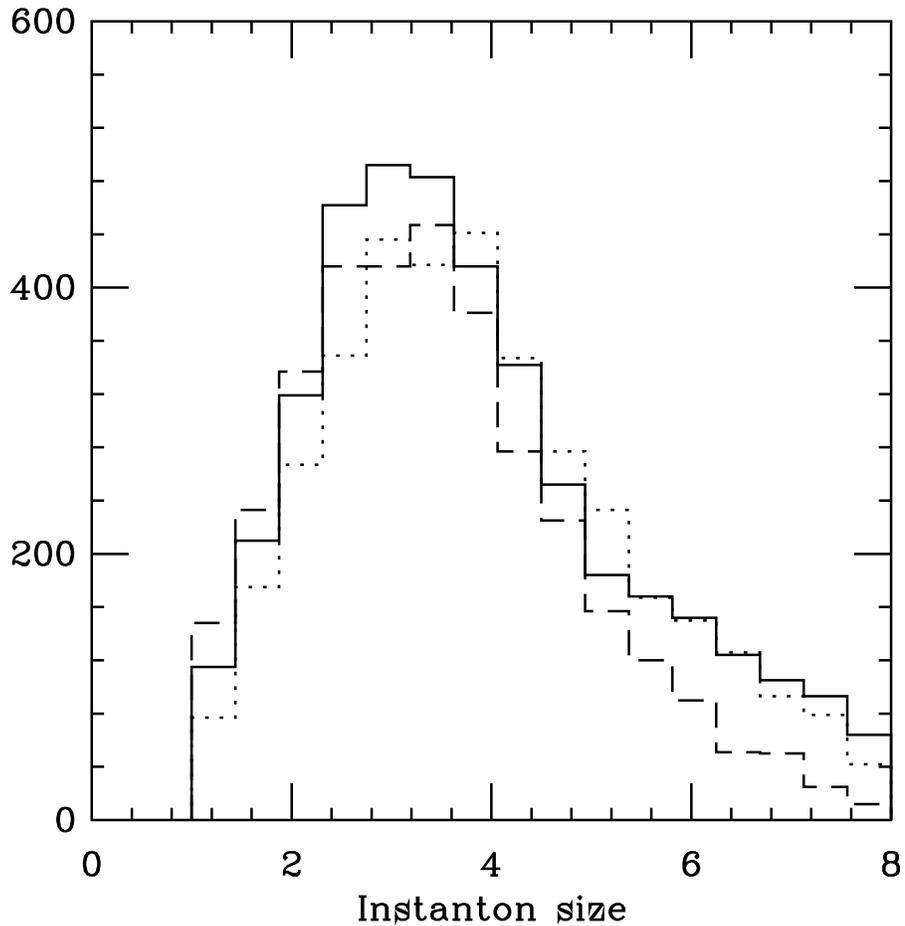}
\end{center}
\caption{The instanton size distribution with the relative 
orientation being distributed according to the Haar measure. The solid 
line indicates the distribution of the ``physical'' (see text) ADHM 
size parameter, the dotted and the dashed lines are the size 
distributions determined by the instanton finder algorithms of 
Ref.\ \protect\cite{I1} and \protect\cite{I2} respectively.}
\label{fig:haar_all}
\end{figure}
In the remainder of the paper we study numerically how the relative
orientation can affect the lattice instanton size distribution. In order 
to see the trends as clearly as possible, instead of trying to mimic the 
``real'' distribution of relative orientations, we use two simple 
orientation distributions: the Haar measure, and all instantons taken 
parallel. Due to the $\sin^2 \! \phi$ factor, the Haar measure very 
strongly favours (close to) perpendicular orientation, thus our two 
distributions almost represent the two possible extremes. We generated 
the charge density of a set of instanton pairs with the ADHM construction 
using eq.\ (\ref{eq:charge}). The parameters of the ADHM ansatz were taken 
as follows. The instanton scale parameters $(\rho_{1,2})$ were distributed 
independently and qualitatively similar to that found on the lattice, 
except for an enhanced tail (for $\rho$ large). We artificially enhanced 
the tail of the distribution in order to test whether such a tail can 
remain undetected by the lattice instanton finders. The separation $2|z|$ 
was Gaussian distributed with mean 7.0, and variance of 1.0.

The resulting charge densities --- each pair resolved on a $16^4$ grid 
--- were then analysed using two different instanton finder algorithms
\cite{I1,I2}. The details of these algorithms are not relevant in the 
present context. However their most important common feature is that 
they are both based on the dilute gas assumption. They identify the 
highest peaks in the charge density and estimate the instanton sizes 
from the fall-off of the density in the vicinity of the maximum.

\begin{figure}[!htb]
\begin{center}
\vskip 10mm
\leavevmode
\epsfxsize=120mm
\epsfbox{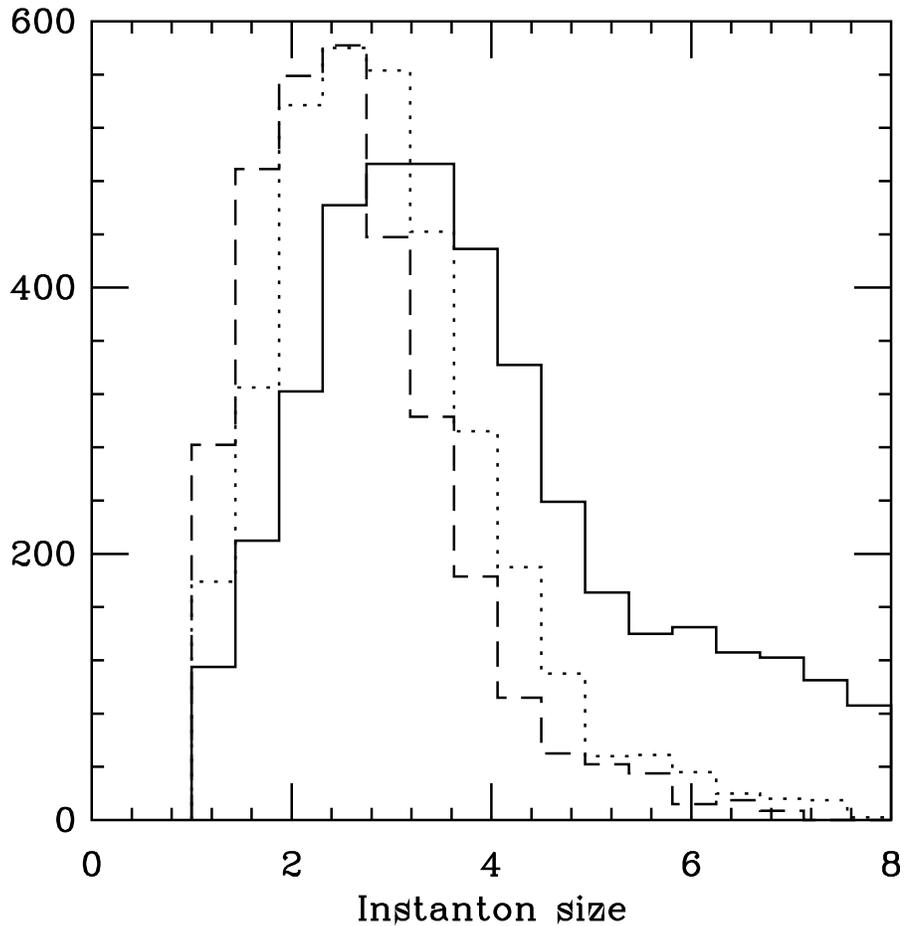}
\end{center}
\caption{The same as Fig.\ \protect\ref{fig:haar_all}, except the relative
orientation in group space is always chosen parallel.}
\label{fig:par_all}
\end{figure}
                               
In Fig.\ \ref{fig:haar_all} we show the instanton size distributions found 
by the algorithms of Refs.\ \cite{I1} (dotted line) and \cite{I2} (dashed 
line) along with the distribution of the ADHM size parameters used to 
construct the charge densities (solid line), representing the physical 
choice of parameters. The Haar measure was used for the gauge orientation.
One of the instanton finders seems to somewhat suppress the enhanced tail 
while this effect is not significant with the other.

In Fig.\ \ref{fig:par_all} we plotted the size distributions obtained when 
all the pairs were taken parallely oriented in group space. All the 
conventions are the same as in Fig.\ \ref{fig:haar_all}. 
Here the two instanton finders both yield a significantly suppressed tail.

We also considered ensembles (not shown) which in terms of the relative 
orientation are in between the two shown. We can draw the following general 
conclusion. The two instanton finders display the same trend; the more 
parallel the pair is in group space the more suppressed the large instantons
become. This is partly due to the effect that large instantons can ``hide'' 
under small ones that produce sharper peaks in the charge density. The other 
effect is that, when two instantons come close with approximately parallel 
orientation, they look narrower\footnote{In the limit of zero separation 
and equal sizes this leads to a singular instanton (on top of the background 
of a large instanton), as discussed in the previous section.} and are thus 
interpreted by the lattice algorithm as ``small'' instantons

\section{Conclusions}

To summarise, we studied the question of what happens when the instanton 
liquid is not dilute enough to be considered as a collection of individual 
pseudoparticles. The next approximation is to treat nearest pairs of like 
charge exactly. We established a correspondence between the parameters of 
individual instantons and the parameters of the charge two ADHM construction. 
After fixing most of the ambiguities of this correspondence we still have 
two dual sets of parameters describing the same charge two instanton 
solution. This duality turned out to imply the existence of a minimal
distance between the two instantons, which is maximal in the case of 
perpendicular orientation. In the other extreme case of (nearly) parallel 
orientation, we found that instanton finders based on the diluteness 
assumption can have a tendency to miss large instantons or to underestimate 
instanton sizes. We numerically confirmed that this indeed happens for 
realistic parameters of the instanton liquid and the effect becomes larger 
for orientations closer to being parallel. In fact, a recent lattice study 
indicates that like charge pairs strongly favour parallel orientation 
compared to the Haar measure \cite{Ilgenfritz}, therefore the effect we 
found can be potentially important.

\section*{Acknowledgements}

We thank Jan de Boer and Thomas Kraan for discussions.
This work was supported in part by a grant from ``Stichting Nationale 
Computer Faciliteiten (NCF)'' for use of the Cray Y-MP C90 at SARA.
T. Kov\'acs was supported by FOM and M. Garc\'{\i}a P\'erez by
CICYT under grant AEN97-1678.

\end{document}